# On the report of discovery of superluminal neutrinos

Moses Fayngold

*Department of Physics, New Jersey Institute of Technology, Newark, NJ 07102*
(Current email address: fayngold@mailaps.org)

So far, none of the known and confirmed experiments have cast any doubts on foundations of special relativity (SR) (see, e.g., experiments on superluminal quantum tunneling [1-3]). The same is true regarding the impact of quantum non-locality; as we know it today, in no way does it challenge SR [4, 5]. In this brief comment we argue that even if the superluminal neutrinos do exist, this fact by itself will not overturn Relativity.

The recent report about the discovery of $\mu$-neutrinos' superluminal propagation in the CERN-OPERA experiment [6] may be based on some subtle systematic error. For instance, it could be another case of observed superluminal group velocity similar to those discussed in [1-3]. This is possible because the earth's crust acts as an optical medium with effective refraction index *n*. Its imaginary part $\text{Im}(n)$ responsible for absorption is definitely dispersive, and so must be the real part $\text{Re}(n)$. This part is extremely close but not exactly equal to 1 for neutrinos and thus can reshape the pulse associated with each particle. It is true that such explanation, if valid, may also apply to interstellar space which is not ideal vacuum, but in this case the known data (e.g. observation of SN 1987A) showed neutrinos to be slightly subluminal [7]. The discrepancy between [6] and [7] may have many different causes and it cannot by itself invalidate the result [6]. Generally, if confirmed, this result will not necessarily call for revision of the existing theory. Quite the contrary, it could consolidate its foundations even further. One can see two reasons for this.

The first, and most important, is that, contrary to the widely spread misconception, SR in no way forbids superluminal motions. The above-mentioned superluminal group velocity in [1-3] is well-known example. Even in the "worst" scenario (if such motions can be harnessed for a signal transfer) they could challenge SR only indirectly through causality violation [8-11]. A possible way to address this problem within the framework of SR has been suggested in [12]. According to [12], it is conceivable that superluminal signaling may lie at the heart of certain kinds of quantum superposition. The mathematical structure of SR explicitly admits the existence of superluminal particles – tachyons [9-14], and moreover, it is *begging* for their existence to satisfy specific symmetry requirements [15, 16]. This symmetry becomes more complete if all known particles and antiparticles have their "tachyonic" counterparts with the same quantum numbers on the other side of the light barrier. Therefore each known particle, neutrino included, may have its superluminal "dual" partner [15, 16] with the same invariant mass $\mu_0$, charge *q*, spin *s*, etc[1].

Different versions of possible superluminal neutrinos were discussed up to now (see, e.g., [17-21].) An interesting possibility of connection between t' Hooft-Polyakov monopoles [22, 23] and superluminal neutrinos has been suggested in [24]. But all these ideas are within the framework of the existing theory.

---

[1] The known models of tachyons based on Quantum Field Theory impose important limitations on their properties. Thus, a model by G. Feinberg [13] describes tachyons as spinless particles obeying the Fermi-Dirac statistics. However, since Feinberg employs the Klein-Gordon equation for tachyons as the underlying assumption, the model cannot be taken as a definitive proof that all tachyons must have zero spin.

The second reason concerns the universal constants. The domain of classical mechanics is restricted by the requirement that classical action $S \gg \hbar$. The non-relativistic physics is restricted by condition $v \ll c$. So far it is not clear which, if any, constant could restrict the domain of SR. It seems unlikely that it could be the same constant *c*, which is the two-sided barrier *within* the framework of SR *and* in the General Relativity as well. In view of the above-mentioned symmetry, one can rather expect an exotic special case of SR in the limit $v \gg c$ (a "mirror image" of the non-relativistic case $v \ll c$); but as far as we know today, there is nothing in SR that would cause its "crash" at $v > c$.

Therefore the result [6] may be the discovery of the very same tachyons that have been unsuccessfully sought for since late 1960-s [25, 26]. One of the reasons of failure in some early attempts to find (electrically charged) tachyons could be the quantum Cerenkov radiation (QCR) causing the uncontrollable deflections from the tachyon's initial direction [27]. This effect must be much less pronounced for electrically neutral particles with the zero electrical dipole moment. In such cases there remains only irreducible gravitational Cerenkov radiation, which is by many orders of magnitude weaker than electromagnetic radiation, especially for the less massive particles. Therefore one could envision that neutrino duals must be the most probable candidates for the first-discovered tachyons.

So if the result [6] is confirmed in the new experiments, this could mean a discovery of the counterparts of neutrino existing on the other side of the light barrier. It would also be an evidence of non-zero interactions between the tachyons and "regular" particles (tardyons), as well as confirmation of non-zero rest mass of the $\mu$-neutrino. In other words, that could be an experimental realization of some previously suggested models of tachyons, including the hypothetical "dual particles" described in [15, 16].

*Summary*:
As any new and potentially important result, the report [6] needs a thorough scrutiny; but if confirmed, it may herald a new triumph of SR rather than its demise.